\documentclass{iopart}

\usepackage{graphics}
\usepackage{epsf}
\usepackage{epsfig}

\begin{document} 


\title[DP and PCP in an external field]{Universal scaling behavior 
of directed percolation and the pair contact process in an external field}

\author{S. L\"ubeck$^{1,2}$ 
and R.\,D.~Willmann$^{1,3}$}

\address{
$^1$Weizmann Institute, Department of Physics of
Complex Systems,\\ 76100 Rehovot, Israel}
\address{
$^2$Theoretische Tieftemperaturphysik, 
Gerhard-Mercator-Universit\"at,\\
47048 Duisburg, Germany 
}
\address{
$^3$Institut f\"ur Festk\"orperforschung,
Forschungszentrum J\"ulich,\\
52425 J\"ulich, Germany \\[2mm]
sven@thp.uni-duisburg.de, r.willmann@fz-juelich.de\\[2mm]
Received 20\,August\,2002
}


{\vspace{-7.0cm}
\noindent
{accepted for publication in \it J.~Phys.~A {\bf}}
\vspace{6.5cm}}


\begin{abstract}
We consider the scaling behavior of
directed percolation and of the pair contact process
with a conjugated field.
In particular we determine numerically the
equation of state and show that both models
are characterized by the same universal scaling
function.
Furthermore we derive the equation of state
for the pair contact process within a mean-field 
approach which again agrees with the mean-field
equation of state of the directed percolation
universality class.
\end{abstract}



\section{Introduction}

In this paper we consider the scaling behavior of
direction percolation (DP) and of the pair contact
process (PCP) numerically in $1+1$ dimensions and 
analytically within
a mean-field approximation (see for a review~\cite{HINRICHSEN_1}).
Both models display a continuous phase transition
from an active to an inactive state.
In contrast to the unique absorbing state of DP
the PCP is characterized by infinitely many
absorbing states.
Despite this difference it was observed in numerical
simulations that the critical exponents describing the
steady state scaling behavior of the PCP 
agree well with those of DP~\cite{JENSEN_2,JENSEN_3}.
This result is in agreement with a renormalization
group analysis of a phenomenological field theory approach
of the PCP which reduces in the steady state to
the corresponding DP equation~\cite{MUNOZ_1}.

In this work we consider the steady state scaling
behavior of DP and the PCP in an external field 
which is conjugated to the order parameter.
In particular we examine the scaling behavior
of the order parameter (equation of state)
as well as of the order parameter fluctuations.
In contrast to previous works on the PCP we 
do not restrict our attention to the critical exponents
but consider the so-called universal scaling functions
too.
In this way we are able to show that both models are
characterized by the same universal scaling behavior,
i.e., the steady state scaling behavior of the PCP
belongs to the universality class of directed percolation.

In the next section we briefly review the scaling
behavior of DP and define the scaling functions.
We consider then in section~\ref{sec:pcp} for the first 
time the PCP in a conjugated field and show that 
the scaling behavior is characterized by the DP critical exponents.
The universal scaling behavior of DP and the PCP
is presented in section~\ref{sec:uni_scaling}. 
The obtained universal equation of state is compared
to the results of a two loop renormalization group
approach of the corresponding Langevin equation~\cite{JANSSEN_2}.
Finally we derive the equation of state of the PCP
within a mean-field approximation.
Therefore we consider the PCP with particle creation
at randomly selected sites.
This random neighbor interaction suppresses long 
range correlations and the model is analytically tractable.

\section{Directed percolation}
\label{sec:dp}

\begin{figure}[b]
  \includegraphics[width=8.6cm,angle=0]{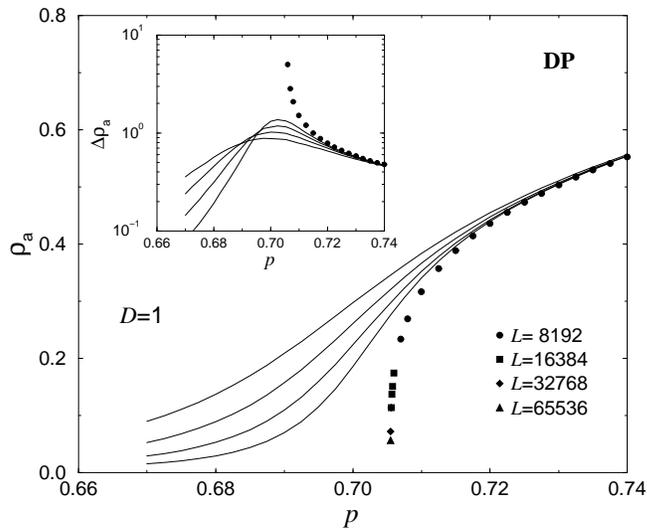}
  \caption{
    The directed percolation 
    order parameter~$\rho_{\rm a}$ as a function of the
    particle density for zero-field (symbols) and for various 
    values of the external field 
    ($h=10^{-4}, 2\,10^{-4}, 5\,10^{-4}, 10^{-3}$) (lines).
    The inset displays the order parameter 
    fluctuations~$\Delta \rho_{\rm a}$ for zero field (symbols)
    and for various values of the external field $h$ (lines). 
   }
  \label{fig:rho_a_1d_dp} 
\end{figure}

In order to examine the scaling behavior of the $1+1$-dimensional 
DP universality class we consider the directed site percolation
process using the Domany-Kinzel automaton~\cite{DOMANY_1}.
It is defined on a diagonal square lattice and evolves 
by parallel update according to the following rules.
A site at time $t$ is occupied with 
probability~$p_{\scriptscriptstyle 2}$ 
($p_{\scriptscriptstyle 1}$) if both (only one) backward
sites (time $t-1$) are occupied.
If both backward sites are empty a 
spontaneous particle creation takes place with 
probability $p_{\scriptscriptstyle 0}$.
Directed site percolation corresponds to the choice
$p_{\scriptscriptstyle 1}=p_{\scriptscriptstyle 2}=p$
and $p_{\scriptscriptstyle 0}=0$.
The propagation probability~$p$ is the control parameter
of the phase transition, i.e., below a 
critical value~$p_{\rm c}$ the activity ceases 
and the system is trapped forever in the absorbing state
(empty lattice).
On the other hand a non-zero density of (active) 
particles $\rho_{\rm a}$ is
found for $p>p_{\rm c}$.
The best estimate of the critical value of directed site
percolation is $p_{\rm c}=0.705489(4)$~\cite{TRETYAKOV_1}
and we use this value throughout this work.

The density of active sites~$\rho_{\rm a}$ is interpreted as the 
order parameter of the absorbing phase transition and 
vanishes at the critical point according to
\begin{equation}
\label{eq:def_beta}
\rho_{\rm{a}} \; \propto \; \delta p^{\beta},
\end{equation}
with $\delta p=(p-p_{\rm c})/p_{\rm c}$.
Furthermore the order parameter fluctuations 
$\Delta\rho_{\rm{a}} = L ( \langle \rho_{\rm{a}}^2 \rangle - \langle
\rho_{\rm{a}}\rangle^2)$  diverge as
\begin{equation}
\label{eq:def_sigma_prime}
\Delta \rho_{\rm{a}} \; \propto \; \delta p^{- \gamma^{\prime}}.
\end{equation} 
The fluctuation exponent $\gamma^{\prime}$ obeys the scaling relation 
$\gamma^{\prime}=\nu_{\scriptscriptstyle \perp}-2 \beta$, 
where $\nu_{\scriptscriptstyle \perp}$ 
describes the divergence of 
the spatial correlation length at the critical point.
The critical behavior of the order parameter is shown in
figure\,\ref{fig:rho_a_1d_dp}.
The data are obtained from numerical simulations of 
systems with periodic boundary conditions.
Considering various system sizes~$L\le 131072$ we 
take care that our results are not affected by finite-size
effects.

\begin{figure}[t]
  \includegraphics[width=8.6cm,angle=0]{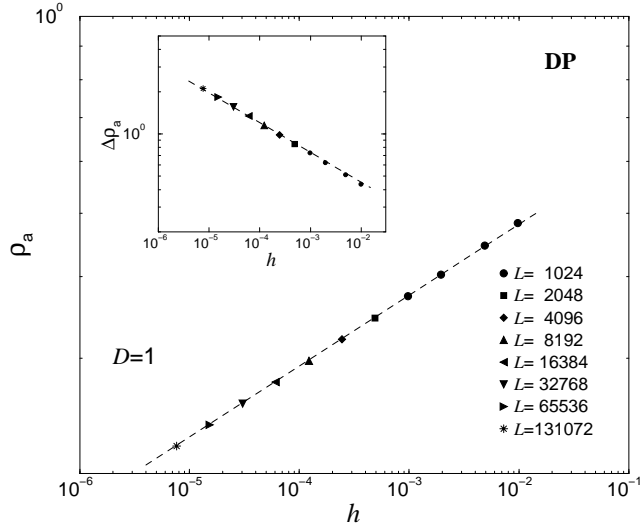}
  \caption{
    The field dependence of the order parameter and its
    fluctuations (inset) at the critical value $p_{\rm c}$
    for directed percolation.
    The dashed lines correspond to the expected power-law
    behavior (\protect\ref{eq:rho_a_field},\protect\ref{eq:scal_fluc}).
   }
  \label{fig:rho_field_1d_dp} 
\end{figure}

Similar to equilibrium phase transitions it is possible
in DP to apply an external field~$h$ that is conjugated
to the order parameter.
Being a conjugated field its has to destroy
the absorbing phase and the corresponding
linear response function has to diverge at the critical
point i.e.,
\begin{equation}
\label{eq:suscept_behavior}
\chi  \; =  \; \frac{\partial\rho_{\rm a}}{\partial h} 
\; \to \; \infty.
\end{equation}
In DP the external field is 
implemented~(see for instance~\cite{HINRICHSEN_1}) as a spontaneous
creation of particles ($p_{\scriptscriptstyle 0}=h>0$).
Clearly the spontaneous creation of particles destroys the
absorbing state and therefore the absorbing phase transition
at all.
Figure\,\ref{fig:rho_a_1d_dp} shows how the external field
results in a smoothening of the zero-field order parameter curve.
The inset displays that the fluctuations are peaked
for finite fields. 
Approaching the transition point ($h\to 0$) this peak
becomes a divergence signalling the critical point.

Close to the transition point the order parameter and its 
fluctuations obey the following scaling ansatzes
\begin{equation}
\label{eq:scal_ansatz_order}
\rho_{\rm{a}}(\delta p , h) \, \sim \,
\lambda\, \, 
{\tilde r}_{\rm \scriptscriptstyle sDP}
(\delta p \; \lambda^{-1/\beta}, h \; \lambda^{-\sigma/\beta}), 
\end{equation}
\begin{equation}
\label{eq:scal_ansatz_fluc}   
\Delta\rho_{\rm{a}}(\delta p , h) \, \sim \,
\lambda^{\gamma^{\prime}}\, \, 
{\tilde d}_{\rm \scriptscriptstyle sDP}
(\delta p \; \lambda, h \, \lambda^{\sigma}) ,
\end{equation}
with the scaling functions ${\tilde r}_{\rm \scriptscriptstyle sDP}$ 
and ${\tilde d}_{\rm \scriptscriptstyle sDP}$  
where the index ${\rm sDP}$ denotes site directed percolation.
The scaling function of the order parameter corresponds to the
equation of state and we recover 
equation~(\ref{eq:def_beta}) by 
setting $\delta \rho \; \lambda^{-1/\beta}=1$ at $h=0$.
On the other hand setting $h \; \lambda^{-\sigma/\beta}=1$ 
leads to 
\begin{equation}
\label{eq:rho_a_field}
\rho_{\rm{a}} (\delta p =0, h) \; \propto \; 
h^{\beta / \sigma}
\end{equation}
at $\delta\rho=0$.
Analogous we get the scaling behavior of the fluctuations 
at $p_{\rm c}$ 
\begin{equation}
\label{eq:scal_fluc}
\Delta \rho_{\rm{a}}(\delta p =0,h) \; \propto \; h^{- \gamma' / \sigma}
\end{equation}
as well as equation~(\ref{eq:def_sigma_prime}).
The field dependence of the order parameter and of its fluctuations 
at $p_{\rm c}$ is plotted in figure~\ref{fig:rho_field_1d_dp}.

\begin{figure}[b]
  \includegraphics[width=8.6cm,angle=0]{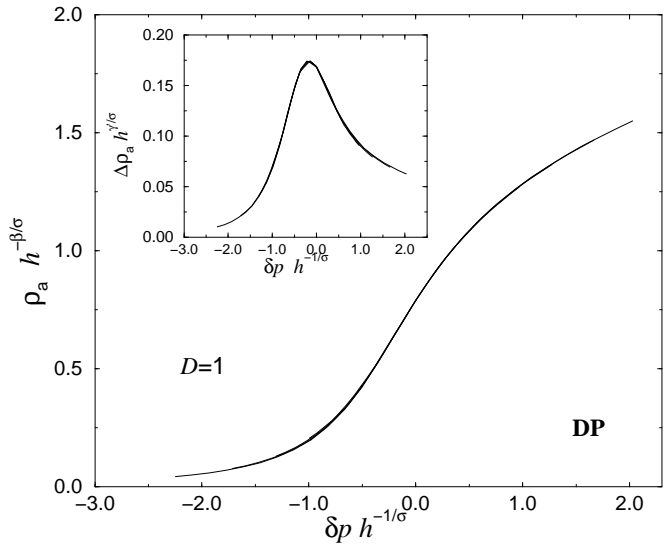}
  \caption{
    The scaling plot of the order parameter and its
    fluctuations (inset) for directed percolation.
   }
  \label{fig:rho_scal_1d_dp} 
\end{figure}

The above scaling forms imply that curves 
corresponding to different values of the external field
collapse to a single one if 
$\rho_{\rm a}\, h^{-\beta/\sigma}$ and 
$\Delta\rho_{\rm a}\, h^{\gamma^{\prime}/\sigma}$ is
considered as a function of the rescaled control parameter
$\delta p \, h^{-1/\sigma}$.
Using the best known estimates of the critical exponents
$\beta=0.276486$,
$\sigma=2.554216$, and
$\gamma^{\prime}= 0.543882$~\cite{JENSEN_5} we get 
beautiful data collapses of our numerical 
data (see figure~\ref{fig:rho_scal_1d_dp}).

It is worth mentioning that
the validity of the scaling behavior of the equation 
of state (\ref{eq:scal_ansatz_order})
implies the required singularity of the
linear response function~(\ref{eq:suscept_behavior}).
Using the fact that the susceptibility
is defined as the derivative of the order parameter with respect to the
conjugated field we get
\begin{eqnarray}
\label{eq:def_suscept}
\chi(\delta p ,h) &  =  & \frac{\partial\hphantom{h}}{\partial h} \,
\rho_{\rm a}(\delta p , h) \\
& \sim &
\lambda^{1-\sigma/\beta}\, \, {\tilde c}_{\rm \scriptscriptstyle sDP}
(\delta p \; \lambda^{-1/\beta}, h \; \lambda^{-\sigma/\beta}).
\end{eqnarray}
Thus we get for the critical behavior 
\begin{equation}
\chi(\delta p ,h)  \; = \;  \left . \frac{\partial\hphantom{h}}{\partial h} \,
\rho_{\rm a}(\delta p, h)
\right |_{h \to 0} \; \propto \; \delta p^{-\gamma}
\label{eq:suscept_gamma_1}
\end{equation}
and 
\begin{equation}
\chi(\delta p,h)  \; = \;  \left . \frac{\partial\hphantom{h}}{\partial h} \,
\rho_{\rm a}(\delta p, h)
\right |_{\delta p \to 0} \; \propto \; h^{-\gamma/\sigma},
\label{eq:suscept_gamma_2}
\end{equation}
respectively.
Here the susceptibility exponent $\gamma$ is given by the
scaling relation
\begin{equation}
\gamma \; = \; \sigma \, - \,  \beta
\label{eq:widom}
\end{equation}
which corresponds to the well known Widom equation  
of equilibrium phase transitions.
Notice that in contrast to the scaling behavior of 
equilibrium phase transitions
the non-equilibrium absorbing phase transition of DP
is characterized by $\gamma \neq \gamma^{\prime}$.

Again, the validity of the scaling ansatz of the order 
parameter (\ref{eq:scal_ansatz_order}) can be used to check
that the implemented field satisfies
the condition (\ref{eq:suscept_behavior}), i.e.,
to check whether the field is conjugated to the order 
parameter or not.
We apply this procedure in the following section
where we examine the scaling behavior of the PCP.

\section{Pair contact process}
\label{sec:pcp}

\begin{figure}[b]
  \includegraphics[width=8.6cm,angle=0]{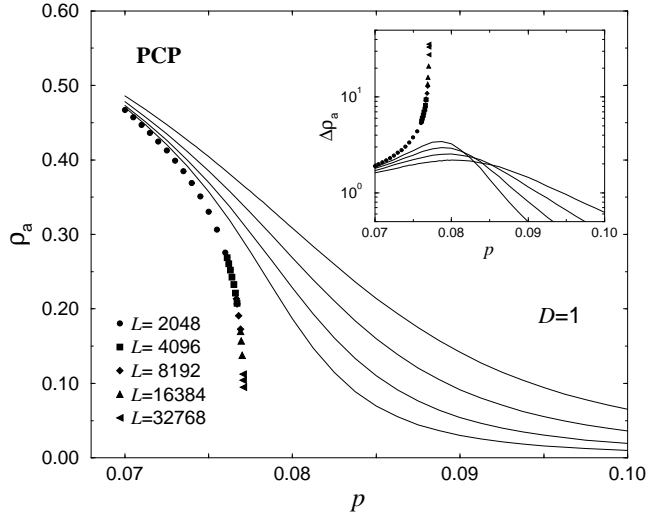}
  \caption{
    The pair contact process
    order parameter~$\rho_{\rm a}$ as a function of the
    particle density for zero-field (symbols) and for 
    various values of the external field from 
    ($h=10^{-4}, 2\,10^{-4}, 5\,10^{-4}, 10^{-3}$) (lines).
    The inset displays the order parameter 
    fluctuations~$\Delta \rho_{\rm a}$ for zero field (symbols)
    and for various values of the external field $h$ (lines). 
   }
  \label{fig:rho_a_1d_pcp} 
\end{figure}

The PCP as introduced by Jensen~\cite{JENSEN_1} is one of the 
simplest models with infinitely
many absorbing states showing a continuous phase transition.
At time $t$ sites on a lattice of length $L$ with 
periodic boundary conditions can
either be occupied ($n_i(t)=1$) or empty ($n_i(t)=0$). 
Pairs of adjacent occupied sites $i,i+1$, linked by an active bond, 
annihilate each other with rate $p$ or create an offspring with rate 
$1-p$ at either site $i-1$ or $i+2$ provided the target site is empty.  
The density of active bonds $\rho_{\rm{a}}$ is the order parameter 
of a continuous phase transition from an active state for $p<p_{\rm c}$ 
to an inactive absorbing state without particle pairs. 
The behavior of the PCP order parameter and its fluctuations
are plotted in figure~\ref{fig:rho_a_1d_pcp}.
The data are obtained from simulations on various system
sizes $L \le 131072$ with periodic boundary conditions.
Our analysis reveals that the critical value is
$p_{\rm c}=0.077093(3)$ which agrees with the 
value $p_{\rm c}=0.077090(5)$~\cite{DICKMAN_4}
obtained from a finite-size scaling analysis of the
lifetime distribution.

In contrast to DP there is no unique absorbing
state (empty lattice) but infinitely many, 
as any configuration with only isolated inactive particles is absorbing. 
Thus in the thermodynamic limit the system will be trapped 
in one of an infinite number of absorbing configurations 
for $p>p_{\rm c}$.
Despite the different structure of the absorbing states the
steady state scaling behavior of the PCP is believed to be
characterized by the 
DP critical exponents $\beta$, $\gamma^{\prime}$, $\gamma$ etc.
On the other hand the dynamical scaling behavior,
associated with activity spreading of a localized seed
depends on the details of the system 
preparation~\cite{JENSEN_3}.

\begin{figure}[t]
  \includegraphics[width=8.6cm,angle=0]{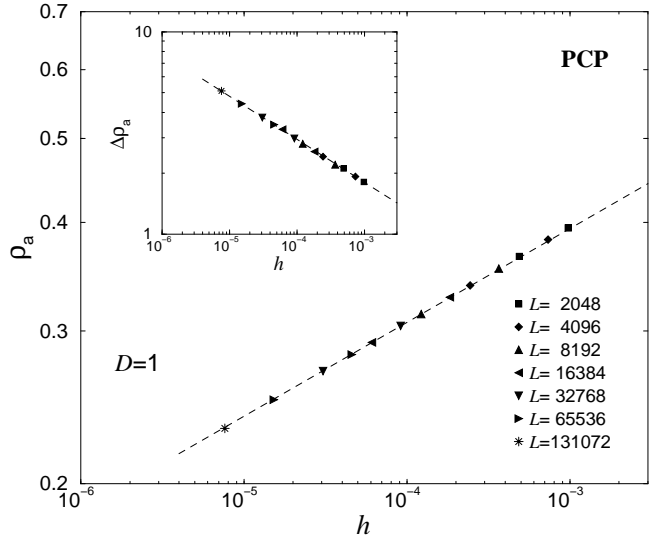}
  \caption{
    The field dependence of the order parameter and its
    fluctuations (inset) at the critical value $p_{\rm c}$
    for the pair contact process.
    The dashed lines correspond to the expected power-law
    behavior (\protect\ref{eq:rho_a_field},\protect\ref{eq:scal_fluc}).
   }
  \label{fig:rho_field_1d_pcp} 
\end{figure}

Recently, Dickman et al.~\cite{DICKMAN_5} considered the PCP with 
an external field that randomly creates isolated particles.
Thus the external field couples to the particle density but not
to the order parameter itself, i.e., the external field is not
conjugated to the order parameter.
The authors observe that the external field shifts the critical 
values $p_{\rm c}$ continuously and that the critical exponents 
are unaffected by the presence of the particle source.
In order to investigate the PCP in a conjugated field 
the implementation of the external field of~\cite{DICKMAN_5}
has to be modified.
Several modifications of the external field are possible.
For instance in absorbing phase transitions with particle 
conservation~\cite{ROSSI_1} the conjugated field triggers movements 
of inactive particles which can be activated in this way~\cite{LUEB_22}.

As shown below spontaneous particle creation with 
rate $h$ acts as a conjugated field analogous to DP.
Figure~\ref{fig:rho_a_1d_pcp} shows how the spontaneous
particle creation smoothens the critical zero field curves
similar to the DP behavior (see figure\,\ref{fig:rho_a_1d_dp}).
We simulated the PCP at the critical value $p_{\rm c}$
for various fields.
The order parameter and its fluctuations as a function
of the external field~$h$ are presented in figure~\ref{fig:rho_field_1d_pcp}.
Approaching the transition point, 
$\rho_{\rm a}$ and $\Delta\rho_{\rm a}$
scale according to the equations (\ref{eq:rho_a_field},\ref{eq:scal_fluc})
where the exponents~$\beta/\sigma$ and $\gamma^{\prime}/\sigma$ 
agree with the DP values.

\begin{figure}[t]
  \includegraphics[width=8.6cm,angle=0]{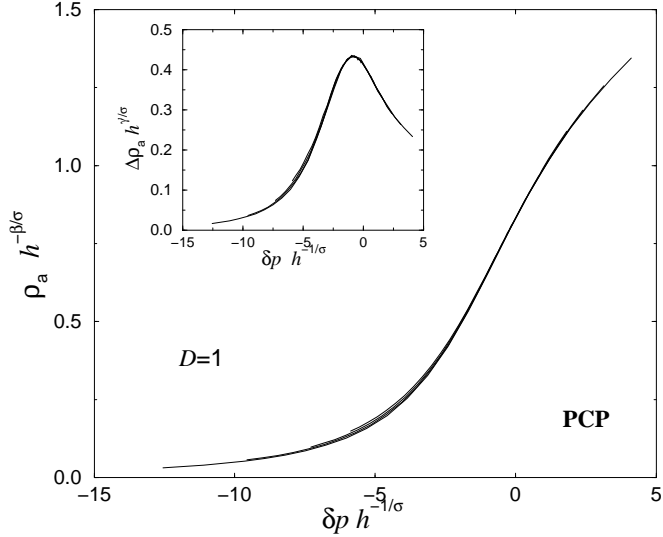}
  \caption{
    The scaling plot of the order parameter and its
    fluctuations (inset) for the pair contact process.
   }
  \label{fig:rho_scal_1d_pcp} 
\end{figure}

Furthermore we assume that the order parameter and the 
order parameter fluctuations obey analogous to DP
the scaling forms
\begin{equation}
\label{eq:scal_ansatz_order_pcp}
\rho_{\rm{a}}(\delta p, h) \, \sim \,
\lambda\, \, 
{\tilde r}_{\rm \scriptscriptstyle PCP}
(\delta p \; \lambda^{-1/\beta}, h \; \lambda^{-\sigma/\beta}) , 
\end{equation}
\begin{equation}
\label{eq:scal_ansatz_fluc_pcp}   
\Delta\rho_{\rm{a}}(\delta p, h) \, \sim \,
\lambda^{\gamma^{\prime}}\, \, 
{\tilde d}_{\rm \scriptscriptstyle PCP}
(\delta p \; \lambda, h \, \lambda^{\sigma}) ,
\end{equation}
where the distance to the critical point is now 
given by $\delta p=(p_c-p)/p_{\rm c}$.
Using the DP values of the critical exponents $\beta$,
$\sigma$ and $\gamma^{\prime}$ we get convincing
data collapses (see figure\,\ref{fig:rho_scal_1d_pcp}).
As pointed out above, the validity of the scaling ansatz
(\ref{eq:scal_ansatz_order_pcp}) implies the singular
behavior of the linear response 
function (\ref{eq:suscept_behavior}), 
i.e., the spontaneous particle
creation in the PCP can be interpreted as an external
field conjugated to the order parameter.
Furthermore the data collapse confirms again that the 
steady state scaling behavior of the PCP 
is characterized by the DP exponents.

\section{Universal scaling behavior}
\label{sec:uni_scaling}

One of the most striking features of critical phenomena
is the concept of universality, i.e., close to the
critical point the scaling behavior depends only on few
fundamental parameters whereas the interaction details
of the systems do not alter the scaling behavior.
In the case of systems with short range interactions
these parameters are the symmetry of the order parameter
and the dimensionality of space~$D$~\cite{GRIFFITHS_1,KADANOFF_2}.
Classical examples of such universal behavior are for instance 
the coexistence curve of liquid-vapor systems~\cite{GUGGENHEIM_1} 
and the equation of state in ferromagnetic systems~\cite{MILOSEVIC_2}.
In the case of absorbing phase transitions the 
corresponding universality hypothesis states 
that systems exhibiting a continuous phase transition 
to a unique absorbing state generally belong to the universality 
class of directed percolation~\cite{JANSSEN_1,GRASSBERGER_2}.
A different scaling behavior is observed only in systems with
additional symmetries (such as parity conservation or particle 
conservation) or in systems with quenched randomness.

\begin{figure}[t]
  \includegraphics[width=8.6cm,angle=0]{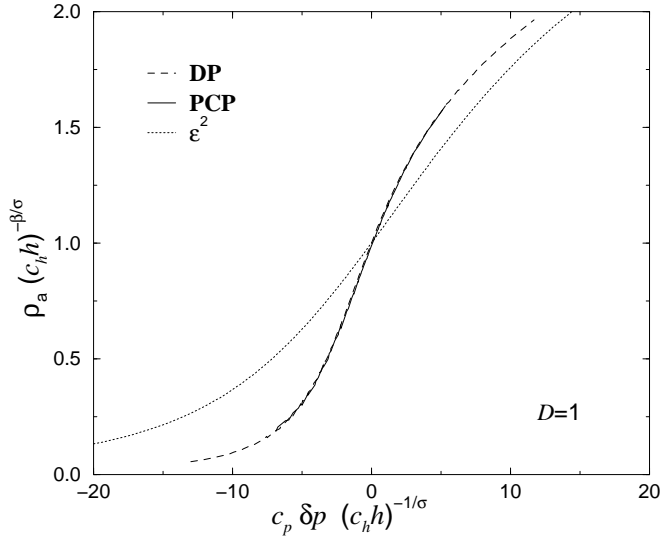}
  \caption{
    The universal order parameter scaling function 
    ${\tilde R}_{\rm \scriptscriptstyle DP}(x,1)$ of the 
    universality class of directed percolation.
    The dotted line corresponds to the result of a
    two loop renormalization group analysis of the
    Langevin equation~\protect\cite{JANSSEN_2} (see text). 
   }
  \label{fig:uni_scal_field_01} 
\end{figure}

Following the concept of universality two models 
belong to the same universality class if the critical exponents 
and the universal scaling functions are identical.
The universal scaling functions 
${\tilde R}_{\rm \scriptscriptstyle DP}$ and 
${\tilde D}_{\rm \scriptscriptstyle DP}$
of the DP universality class 
can be easily determined by introducing
non-universal metric factors $c_i$ and $d_i$ (see e.g.~\cite{PRIVMAN_1}) 
for each scaling argument in the scaling functions
(\ref{eq:scal_ansatz_order},
\ref{eq:scal_ansatz_fluc},
\ref{eq:scal_ansatz_order_pcp},
\ref{eq:scal_ansatz_fluc_pcp}), i.e.,
\begin{equation}
\label{eq:scal_ansatz_order_dp}
\rho_{\rm{a}}(\delta p, h) \, \sim \,
\lambda\;
{\tilde R}_{\rm \scriptscriptstyle DP}
(c_{\scriptscriptstyle 1} \, \delta p \; \lambda^{-1/\beta}, 
c_{\scriptscriptstyle 2} \, h \; \lambda^{-\sigma/\beta}), 
\end{equation}
\begin{equation}
\label{eq:scal_ansatz_fluc_dp}   
\Delta\rho_{\rm{a}}(\delta p, h) \, \sim \,
\lambda^{\gamma^{\prime}}\;
{\tilde D}_{\rm \scriptscriptstyle DP}
(d_{\scriptscriptstyle 1} \, \delta p \, \lambda, 
d_{\scriptscriptstyle 2} \, h \, \lambda^{\sigma}) .
\end{equation}
Using the normalizations 
${\tilde R}_{\rm \scriptscriptstyle DP}(1,0)={\tilde R}_{\rm \scriptscriptstyle DP}(0,1)=1$ 
and 
${\tilde D}_{\rm \scriptscriptstyle DP}(1,0)={\tilde D}_{\rm \scriptscriptstyle DP}(0,1)=1$
the non-universal metric factors can be determined from the amplitudes
of 
\begin{equation}
\label{eq:metric_factors_c}
\rho_{\rm{a}}(\delta p, h=0) \; \sim \; 
(c_{\scriptscriptstyle 1} \, \delta p)^{\beta},
\quad\quad
\rho_{\rm{a}}(\delta p =0, h) \; \sim \; 
(c_{\scriptscriptstyle 2} \, h)^{\beta / \sigma}
\end{equation}
and
\begin{equation}
\label{eq:metric_factors_d}
\Delta \rho_{\rm{a}}(\delta p, h=0) \; 
\sim \; (d_{\scriptscriptstyle 1} \, \delta p)^{- \gamma^{\prime}},
\quad\;
\Delta \rho_{\rm{a}}(\delta p =0,h) 
\; \sim \; (d_{\scriptscriptstyle 2} \, h)^{- \gamma' / \sigma} .
\end{equation}
Like the value of the critical point $p_{\rm c}$ 
the non-universal metric factors may depend on the details
of the considered system, e.g.~the lattice structure, 
the boundary conditions, the used update scheme, etc. 
In the case of directed site percolation in the Domany-Kinzel
automaton we have obtained the values
$c^{\rm \scriptscriptstyle sDP}_{\scriptscriptstyle 1}=2.45$,
$c^{\rm \scriptscriptstyle sDP}_{\scriptscriptstyle 2}=0.112$,
$d^{\rm \scriptscriptstyle sDP}_{\scriptscriptstyle 1}=71.6$ 
and $d^{\rm \scriptscriptstyle sDP}_{\scriptscriptstyle 2}=3984$.
For the PCP on a square lattice we have determined the values
$c^{\rm \scriptscriptstyle PCP}_{\scriptscriptstyle 1}=0.665$,
$c^{\rm \scriptscriptstyle PCP}_{\scriptscriptstyle 2}=0.181$,
$d^{\rm \scriptscriptstyle PCP}_{\scriptscriptstyle 1}=3.21$ 
and $d^{\rm \scriptscriptstyle PCP}_{\scriptscriptstyle 2}=62.11$.

\begin{figure}[t]
  \includegraphics[width=8.6cm,angle=0]{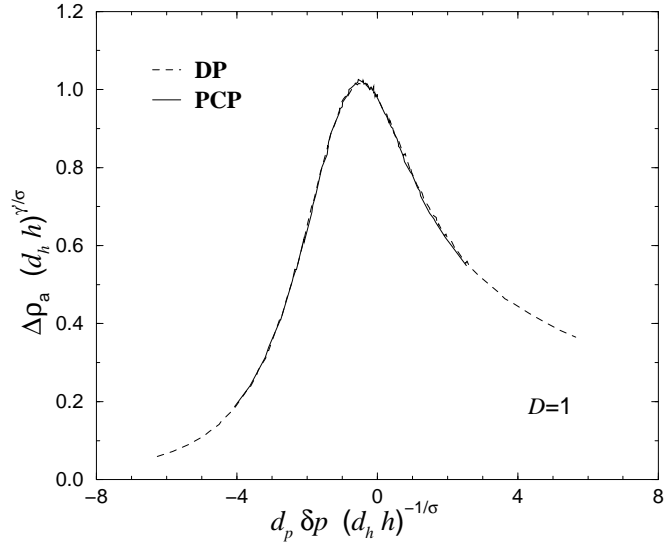}
  \caption{
    The universal scaling function 
    ${\tilde D}_{\rm \scriptscriptstyle DP}(x,1)$ 
    of the order parameter fluctuations of the 
    universality class of directed percolation.
   }
  \label{fig:uni_scal_fluc_01} 
\end{figure}

Analogous to the previous scaling analysis we 
set $c_{\scriptscriptstyle 2}\,h\,\lambda^{-\sigma/\beta}=1$ and 
consider for both models the rescaled order parameter 
$\rho_{\rm a} \, (c_{\scriptscriptstyle 2} h)^{-\beta/\sigma}$
as a function of the rescaled control parameter
$c_{\scriptscriptstyle 1} \delta\rho \, (c_{\scriptscriptstyle 2} h)^{-1/\sigma}$
as well as the rescaled order parameter fluctuations
$\Delta\rho_{\rm a} \, (d_{\scriptscriptstyle 2} h)^{\gamma^{\prime}/\sigma}$
as a function of 
$d_{\scriptscriptstyle 1} \delta\rho \, (d_{\scriptscriptstyle 2} h)^{-1/\sigma}$,
respectively.
The corresponding data are presented in figure\,\ref{fig:uni_scal_field_01}
and figure\,\ref{fig:uni_scal_fluc_01}.
In both cases we get a perfect data collapse of the curves
showing that the one-dimensional PCP steady state scaling
behavior belongs to the universality class of directed 
percolation.

Additionally to the universal scaling 
function ${\tilde R_{\rm \scriptscriptstyle DP}(x,1)}$
we plot in figure\,\ref{fig:uni_scal_field_01} the 
corresponding curve of an $\epsilon$-expansion obtained
from a renormalization group analysis
of a Langevin equation
(see equation~(\ref{eq:langevin_dp}) next section).
Using the parametric representation~\cite{SCHOFIELD_1,JOSEPHSON_1}
of the absorbing
phase transition Janssen {\it et al.} showed recently 
that the equation of state is given by the remarkable 
simple scaling function~\cite{JANSSEN_2} 
\begin{equation}
\label{eq:rg_equation_of_state}
H(\theta) \; = \; \theta \, (2-\theta) + \Or(\epsilon^3),
\end{equation}
where $\epsilon$ denotes the distance to the
upper critical dimension $D_{\rm c}=4$, i.e., 
$\epsilon=D_{\rm c}- D$.
Here the scaling behavior of the 
quantities $\rho_{\rm a}$, $\delta p$, and $h$ is
transformed to the variables $R$ and $\theta$
through the relations
\begin{equation}
\label{eq:para_transform}
b \, \delta p \; = \; R \, (1-\theta),
\quad\quad\quad
\rho_{\rm a} \; = \; R^\beta \, 
\frac{\theta}{2}. 
\end{equation}
The equation of state is given by 
\begin{equation}
\label{eq:para_transform_equa_state}
 a\, h\; = \; 
 \left ( \frac{R^\beta}{2} \right )^{\delta} 
 \; H(\theta) 
\end{equation}
with the metric factors $a$ and $b$.
The whole phase diagram is described by the parameter
range $R\ge 0$ and $\theta \in [ 0 , 2 ]$.
Using the 
equations\,(\ref{eq:rg_equation_of_state},\ref{eq:para_transform},\ref{eq:para_transform_equa_state})
we calculated the corresponding universal function
and compare it in figure~\ref{fig:uni_scal_field_01} 
with the numerically obtained
universal scaling function~${\tilde R_{\rm \scriptscriptstyle DP}(x,1)}$.
As can be seen the significant difference indicates that
the $\Or(\epsilon^3)$ corrections to (\ref{eq:rg_equation_of_state}) 
are relevant, i.e., higher orders than $\Or(\epsilon^2)$ are
necessary to describe the scaling behavior of directed percolation.
Of course we expect that this difference will decrease
with increasing dimension, i.e., for $\epsilon\to 0$.

\begin{figure}[t]
  \includegraphics[width=8.6cm,angle=0]{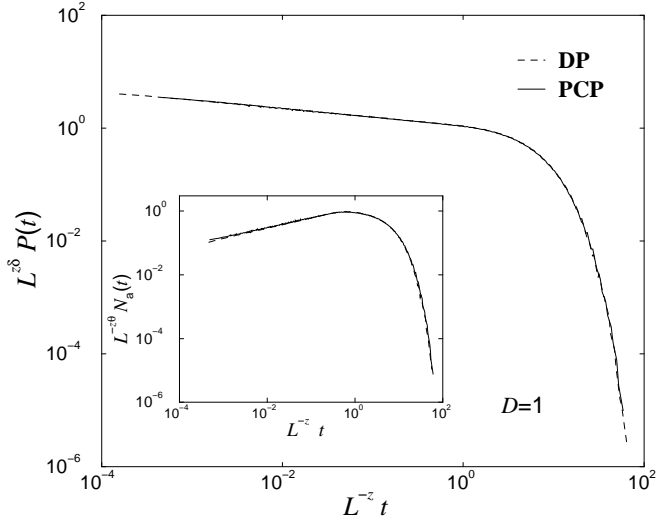}
  \caption{
    The DP universal finite-size scaling function $\tilde P$ 
    and $\tilde N$.
    Both quantities describe the activity spreading of a 
    localized seed (see text).
   }
  \label{fig:p_sur_N_a_scal} 
\end{figure}

Before closing this section we briefly mention that the dynamical
scaling behavior of the PCP belongs to the DP universality
class too if the spreading of a localized
seed is considered at the so-called natural particle density~\cite{JENSEN_3}.
Examining spreading activity one usually considers 
the survival probability $P_{\rm a}$ of the activity as well as 
how the number of active particles~$N_{\rm a}=L \rho_{\rm a}$ 
increases in time.
In the case of DP the simulations are started with a single 
seed on an empty lattice.
For the PCP an absorbing state at $p_{\rm c}$ is prepared
to which a particle is added in order to create one seed 
(one active pair).
At the critical point the following power-law behaviors are expected 
\begin{equation}
\label{eq:spread_quantities}
N_{\rm a} \; \propto  \; t^{\theta} ,
\quad\quad\quad
P_{\rm a} \; \propto  \; t^{-\delta}.
\end{equation}
Finite systems sizes limit these power-law behaviors 
and $P_{\rm a}$ and $N_{\rm a}$ obey 
the finite-size scaling ansatzes
\begin{equation}
\label{eq:spread_quantities_fss_N}
N_{\rm a}(\delta p=0,L) \; \sim  \; \lambda \,
{\tilde n}(\lambda^{-1/\theta} t, \lambda^{-1/\theta z} L)
\end{equation}
\begin{equation}
\label{eq:spread_quantities_fss_P}   
P_{\rm a}(\delta\rho=0,L) \; \sim  \; \lambda \,
{\tilde p}(\lambda^{1/\delta} t, \lambda^{1/\delta z} L).
\end{equation}
where $z$ denotes the dynamical exponent.
Analogous to the above analysis  
the universal scaling curves ${\tilde N}$ and ${\tilde P}$
are obtained by introducing appropriate non-universal metric factors.
Using the values $z=1.580745$, $\theta=0.313686$, 
and $\delta=0.159464$~\cite{JENSEN_5} we get convincing 
data collapses and the universal scaling functions are plotted
in figure~\ref{fig:p_sur_N_a_scal}.

\section{Mean field scaling behavior}
\label{sec:mean_field}

The mean-field equation of state of DP can be easily
derived from the corresponding Langevin equation 
\begin{equation}
\label{eq:langevin_dp}
\partial_t \rho_{\rm{a}} \; = \; 
\delta p \, \rho_{\rm{a}} \, - \, \lambda \rho_{\rm{a}}^2 \, + \,
\kappa \, h \, + \, D \, \nabla^2 \, \rho_{\rm{a}} \, + \, \eta
\end{equation}
which describes the order parameter field $\rho_{\rm a}(\underline x,t)$
on a mesoscopic scale (see~\cite{HINRICHSEN_1}  for a detailed
discussion).
Here $D$ denotes the diffusion constant, $\eta$ denotes
a multiplicative noise term with the correlator
\begin{equation}
\label{eq:langevin_dp_corr}
\langle \, \eta (\underline x,t) \, 
\eta ({\underline x}^{\prime},t^{\prime}) \, \rangle\; = \;
\Gamma \, \rho_{\rm a}(\underline x,t) 
\; \delta^{d}({\underline x}-{\underline x}^{\prime})
\, \delta({t}-{t}^{\prime})
\end{equation}
and $\lambda>0$, $\kappa>0$, and $\Gamma>0$  
are certain coupling constants.
Neglecting spatial correlations and fluctuations ($D=0$ and $\eta=0$)
one gets for the steady state behavior ($\partial_t \rho_{\rm a}=0$)
\begin{equation}
\label{eq:steady_equat}
\delta p \, \rho_{\rm{a}} \, - \, \lambda \, \rho_{\rm{a}}^2 \, 
+ \, \kappa \, h \; = \; 0 
\end{equation}
from which it is easy to derive the universal scaling function
\begin{equation}
\label{eq:mf_scaling}
\rho_{\rm a} \; = \; 
{\tilde R}_{\rm \scriptscriptstyle DP}
(c_{\scriptscriptstyle 1}^{\rm \scriptscriptstyle DP}\, \delta p, 
c_{\scriptscriptstyle 2}^{\rm \scriptscriptstyle DP}
\, h) \; = \;
\frac{c_{\scriptscriptstyle 1}^{\rm \scriptscriptstyle DP} \delta p}{2} \, + \, 
\sqrt{c_{\scriptscriptstyle 2}^{\rm \scriptscriptstyle DP} h \, 
+ \, \left ( 
\frac{c_{\scriptscriptstyle 1}^{\rm \scriptscriptstyle DP} \delta p}{2} 
\right )^2 \;} ,
\end{equation}
where the non-universal metric factors are given by
$c_{\scriptscriptstyle 1}^{\rm \scriptscriptstyle DP}=1/\lambda$ 
and
$c_{\scriptscriptstyle 2}^{\rm \scriptscriptstyle DP}=\kappa / \lambda$,
respectively.
For zero-field we get the solutions $\rho_{\rm a}=0$ (absorbing state)
and $\rho_{\rm a}=c_{\scriptscriptstyle 1}^{\rm \scriptscriptstyle DP} \delta p$, 
i.e., the
mean-field value of the order parameter exponent is $\beta=1$.
On the other hand $\delta p=0$ leads 
to $\rho_{\rm a}= (c_{\scriptscriptstyle 2}^{\rm \scriptscriptstyle DP}\, h)^{1/2}$
implying $\sigma=2$.
As recently shown the universal scaling function (\ref{eq:mf_scaling})
describes not only the scaling behavior of the DP universality
class.
It also occurs in the different universality class of absorbing phase 
transitions with a conserved field~\cite{LUEB_25}.

Let us now consider the following modification of the PCP. 
An active bond produces an offspring with rate $(1-p)$ 
at an empty site selected at random. 
The rules for annihilation and action of the external field 
remain unchanged. 
This random neighbor interaction suppresses long 
range correlations and the model is therefore expected
to be characterized by the mean-field scaling behavior.
We denote the density of inactive bonds between an occupied and an empty 
site as $\rho_{\rm i}$. 
Bonds between empty sites are denoted as  
$\rho_{\rm e}$.
Normalization requires $\rho_{\rm e}+ \rho_{\rm i} + \rho_{\rm a} =1$. 

Depending on the sites adjacent to the target site the number 
of  active bonds $n_{\rm a}$, inactive bonds $n_{\rm i}$ or 
empty bonds $n_{\rm e}$ is changed. 
For instance if the adjacent sites are empty, 
for which the probability in absence of correlations 
is $\rho_{\rm e}^2$, the  number of empty bonds decreases 
by two ($\Delta n_{\rm e}=-2$).
On the other hand there are two new inactive bonds ($\Delta n_{\rm i}=+2$). 
The total probability for this event is $(1-p) \rho_{\rm{a}} \rho_{\rm e}^2$. 
A list of all possible processes and their mean field probabilities is 
given in table~1.
Thus we obtain rate equations for the expectation values $E[\Delta n_x]$ of the 
changes in active, inactive and empty bond numbers. 
These expectation values are zero in the steady state, i.e.,
\begin{equation}
\label{eq:expectation_steady_state}
E[\Delta n_x] \; = \; \sum_{\Delta n_x} \,  \Delta n_x \; p(\Delta n_x) \; = \; 0
\end{equation}
with $x \in \{a,i,e \}$.
In the case of $E[\Delta n_a]$ we get  
\begin{eqnarray}
\label{eq:delta_n_a}
E[\Delta n_a] & = &
-3 p \rho_{\rm{a}}^3 - 4 p \rho_{\rm{a}}^2 \rho_{\rm i} - p \rho_{\rm{a}} \rho_{\rm i}^2 
+2 (1-p) \rho_{\rm{a}} \rho_{\rm i}^2 \\   \nonumber
& & + 2 (1-p) \rho_{\rm{a}} \rho_{\rm i} \rho_{\rm e}
+2h \rho_{\rm i}^2+2h \rho_{\rm i} \rho_{\rm e}
\; = \;0,
\end{eqnarray}
whereas we get for the for inactive bonds
\begin{eqnarray}
\label{eq:delta_n_i}
E[\Delta n_i] & = &
-2 (1-p) \rho_{\rm{a}} \rho_{\rm i}^2 +2 (1-p) \rho_{\rm{a}} \rho_{\rm e}^2 
+ 2p \rho_{\rm{a}}^3\\   \nonumber
& & -2 p \rho_{\rm{a}} \rho_{\rm i}^2 -2h \rho_{\rm i}^2 +2h \rho_{\rm e}^2 
\; =\; 0.
\end{eqnarray}
Using this equations together with the normalization allows for
calculating the order parameter for zero-field ($h=0$) which 
yields the non-trivial ($\rho_{\rm a}>0$) solution 
\begin{equation}
\label{eq:rho_a_mf}
\rho_{\rm{a}} \; = \; 
\frac{8-3 p^2 - 5 p -2 \sqrt{2 p (1-p) (3p-4)^2}\, }{9 p^2-9p+8}.
\end{equation}
This solution is valid below the mean field critical point 
$p_{\rm c}=8/9$ whereas the trivial solution $\rho_{\rm{a}}=0$ 
is valid for all $p$ but unstable above $p_{\rm c}$.
Expanding (\ref{eq:rho_a_mf}) around the critical point
leads to 
\begin{equation}
\label{eq:mf_rho_a_beta}
\rho_{\rm{a}} \; = \; \frac{3}{8} \delta p \, + \, \Or( \delta p^2)
\end{equation} 
with $\delta p=(p_{\rm c}-p)/p_{\rm c}$.
Thus the mean field exponent of the PCP is $\beta=1$ and the 
non-universal metric factor 
$c_{\scriptscriptstyle 1}^{\rm \scriptscriptstyle PCP}=3/8$.

\begin{table}[t]
\caption{The configuration of a PCP lattice before ($\cal{C}$) and
after ($\cal{C'}$) an event.         
Only the sites left and right of those changed by particle creation (top), 
pair annihilation (middle) or particle creation due to the
the external field (bottom) are shown. 
Empty sites are marked by $\circ$, and occupied sites by $\bullet$.
Here, $\Delta n_a$ denotes the change of the number of active
bonds, $\Delta n_i$ the respective change of inactive bonds,
$\Delta n_e$ that of empty bonds
and $P$ is the
corresponding probability of the event
if spatial correlations are neglected.}
\label{table:pcp_conf}
\begin{indented}
\item[]
\begin{tabular}{ccrrrl}
\br
$\cal{C}$ & $\cal{C'}$       & $\Delta n_a$  & $\Delta n_i$  & $\Delta n_e $   & $p(\cal{C}
\to \cal{C}^\prime)$ \cr  \mr
$ \bullet \circ \bullet $     &  $ \bullet \bullet \bullet $ & +2 & $ -2 $ & 0 & $ (1-p) \rho_{\rm{a}} \rho_{\rm i}^2 $ \cr
$ \bullet \circ \circ $       &  $ \bullet \bullet \circ $ & +1 & $ 0 $ & -1 & $ (1-p) \rho_{\rm{a}} \rho_{\rm i} \rho_{\rm e} $ \cr
$ \circ \circ \bullet $     &  $ \circ \bullet \bullet $ & +1 & $ 0 $ & -1 & $ (1-p) \rho_{\rm{a}} \rho_{\rm i} \rho_{\rm e} $ \cr
$ \circ \circ \circ $     &  $ \circ \bullet \circ $ & 0 & $ +2 $ & -2 & $ (1-p) \rho_{\rm{a}} \rho_{\rm e}^2 $ \cr
\mr
$ \bullet \bullet \bullet \, \bullet $ &  $ \bullet \circ \circ \, \bullet $ & -3 & $ +2 $ & +1 & $ p \rho_{\rm{a}}^3 $ \cr
$ \circ \bullet \bullet \, \bullet $ &  $ \circ \circ \circ \, \bullet $ & -2 & $ 0 $ & +2 & $ p \rho_{\rm{a}}^2 \rho_{\rm i}$ \cr
$ \bullet \bullet \bullet \, \circ $ &  $ \bullet \circ \circ \, \circ $ & -2 & $ 0 $ & +2 & $ p \rho_{\rm{a}}^2 \rho_{\rm i}$ \cr
$ \circ \bullet \bullet \, \circ $ &  $ \circ \circ \circ \, \circ $ & -1 & $ -2 $ & +3 & $ p \rho_{\rm{a}} \rho_{\rm i}^2$ \cr
\mr
$ \bullet \circ \bullet $     &  $ \bullet \bullet \bullet $ & +2 & $ -2 $ & 0 & $ h\rho_{\rm i}^2 $ \cr
$ \bullet \circ \circ $       &  $ \bullet \bullet \circ $ & +1 & $ 0 $ & -1 & $ h \rho_{\rm i} \rho_{\rm e} $ \cr
$ \circ \circ \bullet $     &  $ \circ \bullet \bullet $ & +1 & $ 0 $ & -1 & $ h \rho_{\rm i} \rho_{\rm e} $ \cr
$ \circ \circ \circ $     &  $ \circ \bullet \circ $ & 0 & $ +2 $ & -2 & $ h \rho_{\rm e}^2 $ \cr
\br
\end{tabular}
\end{indented}
\end{table}

In order to obtain the order parameter in presence of an external 
field~$h$ equations (\ref{eq:delta_n_a},\ref{eq:delta_n_i}) are solved
for $\rho_{\rm i}$ which yields
\begin{eqnarray}
\label{eq:rho_i_equated}
4h + 4 \rho_{\rm a} -4 h \rho_{\rm a} -4p \rho_{\rm a} -4 \rho_{\rm a}^2 & &\\ \nonumber
+\left \{
{-12 p^2 \rho_{\rm a}^2 + (2h + 2 \rho_{\rm a} -2h \rho_{\rm a} -2p \rho_{\rm a}
-2 \rho_{\rm a}^2 -2p \rho_{\rm a}^2)^2} \right \}^{1/2} \\ \nonumber
-\left \{ {(-2h -2 \rho_{\rm a} +2p \rho_{\rm a} 
+2 \rho_{\rm a}^2 -2p \rho_{\rm a}^2)^2 + \ldots \; } \right . & &\\ \nonumber
\left . {\;\ldots +4 p \rho_{\rm a}(h+\rho_{\rm a}-2h \rho_{\rm a} -p \rho_{\rm a} -2
\rho_{\rm a}^2 +h \rho_{\rm a}^2 +2 p \rho_{\rm a}^2 + \rho_{\rm a}^3)\;}
\right \}^{1/2} &= &0.
\end{eqnarray}
To obtain the field dependence of the order parameter a series expansion 
around $h=0$ at $p_{\rm c}=8/9$ is performed which results 
in leading order
\begin{equation}
\label{eq:rho_a_field_pcp}
\rho_{\rm a} \; = \; \sqrt{\frac{3}{8} \,h },
\end{equation}
i.e., the mean field values of the PCP are given by
$c_{\scriptscriptstyle 2}^{\rm \scriptscriptstyle PCP}=\sqrt{3/8}$ and $\sigma=2$.

Finally we derive the mean-field universal scaling function ${\tilde R}$ 
of the PCP.
Therefore we write (\ref{eq:rho_i_equated})
as a function of the reduced control parameter~$\delta p$   
and perform the limits $\rho_{\rm a}\to 0$, $\delta\rho\to 0$ and $h\to 0$
with the constraint that $\rho_{\rm a}/\sqrt{h}$ and $\rho_{\rm a}/\delta p$ 
are finite. 
Thus we remains in leading order with 
\begin{equation}
\label{eq:pcp_mf}
\frac{3}{8} \, \delta p \, \rho_{\rm a} \, - \,
\rho_{\rm a}^2 \, + \,\frac{3}{8}\, h \; = \;0 .
\end{equation}
Solving this equation yields
\begin{equation}
\label{eq:eq:pcp_mf_uni}
\rho_{\rm a} \; = \; \frac{1}{2} \, \frac{3}{8} \, \delta p 
\, + \, \sqrt{ \frac{3}{8}\, h \, + \, \left (
\frac{1}{2} \, \frac{3}{8}\, \delta p \right )^2 \;}
\; = \; 
{\tilde R}_{\rm \scriptscriptstyle DP}
(c_{\scriptscriptstyle 1}^{\rm \scriptscriptstyle PCP}\, \delta p, 
c_{\scriptscriptstyle 2}^{\rm \scriptscriptstyle PCP} \, h).
\end{equation}
Thus we have shown that both models are characterized by the same 
critical exponents and scaling function, i.e., 
similar to the $1+1$-dimensional case the mean-field 
solution of the PCP belongs to the mean-field universality
class of directed percolation.\\[5mm]

We would like to thank H.~Hinrichsen for usefull discussions.
This work was financially supported by the 
Minerva Foundation (Max Planck Gesellschaft).\\[1cm]

\end{document}